# Assessing and improving an approach to delay-tolerant networking

Lloyd Wood

*Research Fellow, Centre for Communication Systems Research at the University of Surrey, e-mail: L.Wood@surrey.ac.uk*

Abstract

Delay-tolerant networking (DTN) is a term invented to describe and encompass all types of long-delay, disconnected, disrupted or intermittently-connected networks, where mobility and outages or scheduled contacts may be experienced. 'DTN' is also used to refer to the Bundle Protocol, which has been proposed as the one unifying solution for disparate DTN networking scenarios, after originally being designed solely for use in deep space for the 'Interplanetary Internet.' We evaluated the Bundle Protocol by testing it in space and on the ground. We have found architectural weaknesses in the Bundle Protocol that may prevent engineering deployment of this protocol in realistic delay-tolerant networking scenarios, and have proposed approaches to address these weaknesses.

## 1. Introduction

Delay-tolerant networking was originally proposed as a generalisation of NASA-led work to move to packetized networking for its spacecraft. That work was named the 'Interplanetary Internet.' There, long propagation delays between spacecraft and scheduled, planned, contacts dominate communications. Extending the scope of the problem space to also include addressing very different, disrupted, terrestrial ad-hoc networks, including military networks, significantly increased interest in and funding for this new approach to networking, and has led to further development of the Bundle Protocol suite. The Bundle Protocol attempts to encompass many environments and use cases beyond its original deep space scenario, even though those other cases can be very different in connectivity and networking requirements [fig. 1].

We completed the first in-space tests of the Bundle Protocol for the Interplanetary Internet on the UK-DMC satellite [1], and have combined our practical experience with theoretical analysis to provide a detailed consideration of many technical aspects of the Bundle Protocol. The design of the Bundle Protocol ignores the reliability concerns that led to the development of the well-known 'end-to-end principle,' [2] and also raises other technical issues. The issues that we have uncovered include the important reliability and timing problems that we highlight here [3].

## 2. Technical Approach

The Bundle Protocol is intended to embody a new architectural approach to networking. It is not by itself directly compatible with other networking protocols such as the Internet Protocol suite, and cannot be as it attempts to introduce new approaches to identification, addressing and routing. However, the Bundle Protocol can be layered over these other networks using 'convergence layer adapters' [fig. 2]. Given the prevalence of IP networking, most Bundle Protocol development has been with convergence layer adapters for the existing TCP/IP suite, although there has been some work over other protocol suites, including CCSDS for space agencies and AX.25 for ham radio use.

The Bundle Protocol specifies a new way of transmitting data in a complex protocol format that is assembled from different blocks for different purposes. Blocks and header information can be inserted, removed and modified by intermediate nodes.

Emphasis on security has been a focus of the design of the Bundle Protocol from an early stage, with a complex security architecture that provides authentication of messages and encryption of data delivered. This is a deliberate change from earlier architectures; security was famously deliberately left out of the Internet's TCP/IP suite, and had to be retrofitted later with IPsec, HAIPE, SSL and other protocols. However, this focus on security has neglected protocol reliability.

Transmissions and memory storage do not always produce perfect copies (although we may wish to believe so) and do have non-zero error rates. Any introduced errors must be detected with deliberate checks. A well-designed network protocol will sanity-check its headers to make sure that the information it is exchanging was received reliably before being processed. It may also sanity-check its payload data. Checking payload data must also be done in any case by the highest networking layer handling the payload, in accordance with the end-to-end principle, to detect introduced errors.

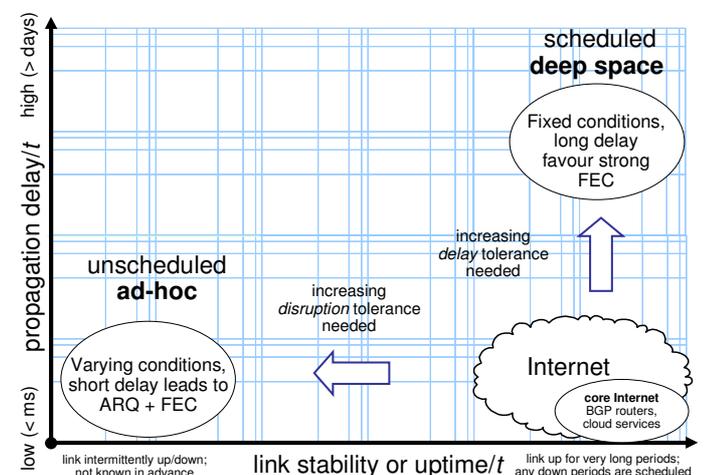

**Figure 1 – Comparing different DTN scenarios**

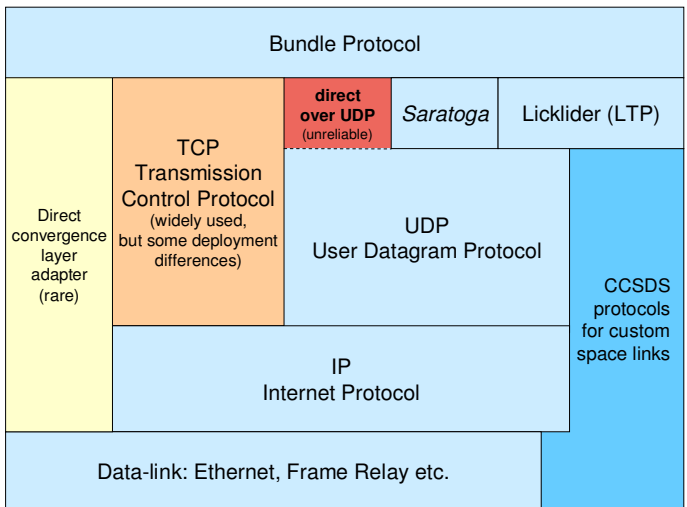

**Figure 2 – Convergence layers under the Bundle Protocol**

In practice, responsibility for end-to-end reliability usually falls to the protocol providing end-to-end transport, which is trusted by local applications – here, that is the Bundle Protocol. When payload data is encrypted or authenticated, a reliability check comes as a side-effect of the security check.

However, the ability to permit block information to be deliberately altered *en route* in the Bundle Protocol, without security checks on that in-the-clear information, because any detectd alteration is viewed as an attack, weakens overall reliability. Errors can be introduced and can go undetected.

The Bundle Protocol also includes node-to-node authentication. This can provide a lower-level reliability check, again as a side-effect of a security mechanism. Alternatively, the Bundle Protocol can rely on the reliability checks in convergence layers and lower protocols. In either case, relying on layers underneath the Bundle Protocol to guarantee correctness of data sent by the Bundle Protocol is hoping for the best in violation of the end-to-end principle, and leads to a more complex protocol stack.

The approach adopted by the Bundle Protocol requires high-complexity, processor-intensive, security mechanisms to be implemented just to provide an approximation of functionality of a lightweight checksum, as a side-effect of the authentication and encryption that the security mechanisms provide. The threat model for the environment may not require the level of security offered by the security architecture, but will require end-to-end reliability checks in accordance with the end-to-end principle. As a result, the security mechanisms are now required to be implemented to gain an assurance of end-to-end reliability. This is an added cost to deploying the Bundle Protocol.

## 3. Results

Known deployments of the Bundle Protocol have run without any security being implemented. Three in-space tests of the protocol for the Interplanetary Internet – in our UK-DMC satellite tests and later on the Deep Impact/EPOXI comet probe [1] and on the International Space Station – chose not to implement bundle security. Not doing so can be attributed to a number of different reasons, including reliance on lower layers for 'probably good enough' reliability, lack of security code and specification readiness, lack of available memory to store and run code, lack of any threat to be worth mitigating against, and security not being required to be able to demonstrate the protocol running in space. The complexity of the Bundle Protocol is one argument against placing it in low-end embedded systems, and processing hardware for space is often low-end and unable to execute modern cryptographic algorithms rapidly. Non-essential processing is not done.

Recent EU trials in a remote area of Sweden also did not implement bundle security, and so are also running without high-level transport reliability checks as a side-effect of not having security [4]. The risks to data of doing so are well-known, and are described in the end-to-end literature [2].

The design of the Bundle Protocol is such that we suggest adding support for lightweight reliability checking within the imposed limits of the existing security framework [5]. Unfortunately, that workaround uses the complex security architecture and requires it to be implemented, so this is unlikely to see widespread adoption in embedded systems.

The Bundle Protocol also expects that all communicating nodes have a shared understanding of UTC time and its leap seconds, with synchronised clocks. Bundles expire after a set clock time and are discarded. Bundles sent from nodes with misset or drifting clocks may be expired at the next node simply because their timestamps are in the far past or distant future. If you don't know the time, you can't ask for the time by using the Bundle Protocol. A bundle age extension block has since been proposed for when UTC time is not known, but setting the age still needs working, reliable, clocks.

## 4. Summary of the work, potential impact and conclusion

We have evaluated the Bundle Protocol, highlighted architectural problems in its design, and proposed a workaround to give reliability. Our work shows that the basic design of the Bundle Protocol neglects important architectural issues. We expect this to limit its adoption and deployment.

## References

[1] W. Ivancic, W. M. Eddy, D. Stewart, L. Wood, J. Northam and C. Jackson, 'Experience with delay-tolerant networking from orbit,' International Journal of Satellite Communications and Networking, vol. 28, issues 5-6, pp. 335-351, Sep.-Dec. 2010.

[2] J. Saltzer, D. Reed, and D. D. Clark, 'End-to-End Arguments in System Design,' ACM Transactions on Computer Systems, 2(4), pages 277-288, 1984.

[3] L. Wood, W. M. Eddy and P. Holliday, 'A Bundle of Problems,' IEEE Aerospace Conference, Big Sky, Montana, March 2009.

[4] S. Farrell, 'Security in the Wild,' IEEE Internet Computing, vol. 15 issue 3, pp. 86-91, May-June 2011.

[5] W. M. Eddy, L. Wood and W. Ivancic, 'Reliability-only Ciphersuites for the Bundle Protocol,' work in progress as an internet draft, adopted as a workgroup draft by the IRTF Delay-Tolerant Networking Research Group, May 2011.